\documentclass[12pt]{article}
\normalbaselines \topmargin -0.25 truein \oddsidemargin 0.30
truein \evensidemargin 0.30 truein 

\font\tbf = cmbx12

\begin{document}

%%%%%%%%%%%%%%%%%%%%%%%%%%%%%%%%%%%%%%%%%%%%%%%%%%%%%%%%%%%%%%
\indent \vskip 1cm
\begin{center}
{\bf NON-MINIMAL PP-WAVE EINSTEIN-YANG-MILLS-HIGGS MODEL: COLOR
CROSS-EFFECTS INDUCED BY CURVATURE}

\end{center}
\vskip 1 cm

%%%%%%%%%%%%%%%%%%%%%%%%%%%%%%%%%%%%%%%%%%%%%%%%%%%%%%%%%%%%%%%
\centerline{\tbf Alexander Balakin\footnote{e-mail:
Alexander.Balakin@ksu.ru},  }

\vskip 0.3cm

\centerline{\it Kazan State University,}

\centerline{\it Kremlevskaya street 18, 420008, Kazan, Russia,}

\vskip 0.3cm

\centerline{\tbf Heinz Dehnen\footnote{e-mail:
Heinz.Dehnen@uni-konstanz.de}, } \vskip 0.3cm

\centerline{\it Universit\"at Konstanz, Fachbereich Physik,}
\centerline{\it Fach M677, D-78457, Konstanz, Germany,}

\vskip 0.3cm \centerline{\tbf and}

\vskip 0.3cm

 \centerline{\tbf  Alexei Zayats\footnote{e-mail: Alexei.Zayats@ksu.ru} } \vskip 0.3cm

\centerline{\it Kazan State University,}

\centerline{\it Kremlevskaya street 18, 420008, Kazan, Russia}

\vskip 0.3cm

\begin{abstract}
Non-minimal interactions in the pp-wave Einstein - Yang - Mills - Higgs
(EYMH) model are shown to give rise to color cross-effects
analogous to the magneto-electricity in the Maxwell theory. In
order to illustrate the significance of these color cross-effects,
we reconstruct the effective (associated, color and
color-acoustic) metrics for the pp-wave non-minimal
seven-parameter EYMH model with parallel gauge and scalar
background fields. Then these metrics are used as hints for
obtaining explicit exact solutions of the non-minimally extended
Yang-Mills and Higgs equations for the test fields propagating in
the vacuum interacting with curvature. The influence of the
non-minimal coupling on the test particle motion is interpreted in
terms of the so-called trapped surfaces, introduced in the Analog
Gravity theory.

\end{abstract}

\vspace{1.5cm}

\newpage

\section{Introduction}

Magneto-electricity is the well-known linear cross-effect
appearing in (moving) aniso\-tropic media \cite{O,HehlObukhov}.
The interest to this problem has been recently revived due to
discussions about the Feigel effect (see \cite{F1,HO21,HO22,HO23}
and references therein), which takes place in crossed external
electric and magnetic field. In this connection we have to stress
that the non-minimal interactions (induced by curvature) also
produce magneto-electric effect in the $U(1)$ symmetric
electrodynamic systems  and color cross-effects in the $SU(N)$
symmetric Yang-Mills-Higgs systems. The signal about
cross-effects appear when we analyze the linear response tensor
in the non-minimal Einstein-Maxwell theory (see, e.g.,
\cite{B1,B2,B4}) and non-minimal Einstein-Yang-Mills-Higgs theory
(see, e.g., \cite{BDZ}). When a medium possesses cross-effects
the theory of wave propagation becomes much more sophisticated.
In order to clarify basic properties of such processes, a
formalism of effective metrics is being widely used (see, e.g.,
\cite{HehlObukhov,Visser1,Novello,VolovikBook}). The development
of this formalism exhibits few interesting details. When we deal
with propagation of electromagnetic waves in vacuum, the inverse
spacetime metric $g^{ik}$ can be regarded both: as an {\it
optical} metric and as an {\it associated} metric. The first term
means, that the photon with the momentum four-vector $p_k$ moves
in pure vacuum along the null geodesic line of this spacetime,
and the eikonal equation $g^{ik}p_i p_k=0$ is satisfied. The
second term, associated metric, means that the linear
constitutive equation $H^{ik}=C^{ikmn}F_{mn}$, linking the
Maxwell tensor $F_{mn}$, the strength of the electromagnetic
field, with the induction tensor
$H^{ik}$\cite{MauginJMP,HehlObukhov}, can be reconstructed for
the vacuum case if we put $C^{ikmn}=
1/2(g^{im}g^{kn}-g^{in}g^{km})$, i.e., using the quadratic
combinations of the inverse metric $g^{ik}$. For the spatially
isotropic medium Gordon \cite{Gordon} introduced the optical
metric $g^{*ik}= g^{ik} + (n^2-1) U^iU^k$ (we use the rule
$U^kU_k=1$ for the normalization of the velocity four-vector
$U^k$). The propagation of photons in such spatially isotropic
medium with a refraction index $n$ is equivalent to the motion in
the effective spacetime with metric $g^{*ik}$, the eikonal
equation $g^{*ik}p_ip_k=0$ being valid. On the other hand, the
tensor of linear response $C^{ikmn}$ can be reconstructed as one
for the vacuum, if to replace $g^{ik}$ by the $g^{*ik}$
\cite{PMQ,MauginJMP}. Thus, the metric $g^{*ik}$ is both optical
and associated one for the spatially isotropic medium.

When the medium is anisotropic, but possesses uniaxial symmetry and cross
magneto-electric effects are absent, then
the quartic Fresnel surface \cite{HehlObukhov} is factorized into the
product of two light cones, and {\it two} optical
metrics of the Lorentzian type are sufficient to describe the
photon propagation \cite{Perlick}. These optical metrics can be
constructed by using not only the terms $g^{ik}$ and $U^iU^k$, but
$X^iX^k$ also, where $X^k$ is a space-like four-vector pointing
the direction of anisotropy in the space. Such bimetricity
relates to the birefringence properties of the anisotropic media.
In its turn, the tensor of linear response $C^{ikmn}$ for the
uniaxial medium without cross-effects can also be reconstructed in
terms of quadratic combinations of two optical metrics
\cite{HehlObukhov,GRG05}, but the corresponding decomposition is
more complicated than the one for the spatially isotropic medium.
When the anisotropic medium is biaxial, Perlick \cite{Perlick}
assumes that the Lorentzian optical metrics do not exist, and one
need to use the optical metrics of the Finsler type, i.e.,
depending not only on the coordinates, but on the particle
four-vector of momentum also.

When a medium possesses cross magneto-electric effects,
the quartic Fresnel surface  can not be
factorized, in general, into the product of two light cones (\cite{HO21,HO23}). However
it becomes possible, when the magneto-electric coefficients obey some special requirements,
and the authors of \cite{HO23} gave an example of such explicit reconstruction of
the two optical metrics.
In order to find associated metrics for the medium, possessing cross magneto-electric
effects, we focus on the formalism proposed in \cite{GRG05}.
This generalized formalism is based
on the introduction of a new effective space, in which the
symmetric tensor fields of the rank two (the associated metrics),
which we use for the reconstruction of the tensor $C^{ikmn}$, are
considered as vectors. When $C^{ikmn}$ is symmetric  with respect
to the transposition of the pairs of indices $ik$ and $mn$, this
effective space is four-dimensional, since arbitrary linear
response tensor can always be decomposed into the sum of four
linearly independent terms $X^i_{(a)}X^k_{(a)}$
($(a)=(0),(1),(2),(3)$), where $X^k_{(a)}$ are tetrad four-vectors.
The representation of the linear response tensor by the
associated metrics is not unique, thus, we considered in
\cite{GRG05} transformations in this effective space related to
the transition from one set of associated metrics to another one.
Based on this idea, we can conclude, that when we deal with
vacuum and spatially isotropic medium, it is sufficient to use
one-dimensional sub-space of this effective space, since there
exists a unique optical metric, reconstructing the tensor
$C^{ikmn}$. In the uniaxial medium without cross-effects it is
sufficient to use two-dimensional sub-space, since two optical
metrics reconstruct the constitutive equations. As for the
biaxial case we need three or four associated metrics, thus, the
number of optical metrics (again two, since the photons have two
degrees of freedom) does not coincide with the number of the
associated ones.

In order to illustrate this idea, we consider here the associated
metrics and color metrics (generalization of the optical metrics
for the Yang-Mills field) for the pp-wave non-minimal
Einstein-Yang-Mills-Higgs (EYMH) model. The symmetry of this model
is not uniaxial, the cross-effect is present, nevertheless, color
metrics and linear response tensor can be reconstructed
explicitly.

The paper is organized as follows. In Section 2 we introduce
briefly the action functional and master equations for the gauge,
scalar and gravitational fields in the framework of
seven-parameter non-minimal EYMH model. In Section 3 we apply
this non-minimal EYMH model to the field configuration with
pp-wave symmetry and reconstruct, subsequently, associated, color
and color-acoustic metrics appearing in the pp-wave non-minimal
EYMH model. In Sections 4 and 5, using the obtained color and
color-acoustic metrics as hints, we obtain the exact parallel in
the group space solutions to the Yang-Mills and Higgs equations,
respectively. In Discussion we interpret the obtained results in
terms of trapped surfaces, introduced in the theory of Analog
Gravity.

\section{Seven-parameter non-minimal EYMH model}

\subsection{Basic definitions}

Consider an action functional
\begin{eqnarray}\label{1act}
S_{({\rm NMEYMH})} = \int d^4 x \sqrt{-g}\ \left\{ \frac{R + 2
\Lambda}{\kappa}+\frac{1}{2}F^{(a)}_{ik} F^{ik}_{(a)}
-{D}_m\Phi^{(a)}{D}^m\Phi_{(a)}+ m^2 \Phi^2+ \right.
{}\nonumber\\
\left. {}+\frac{1}{2} {\cal R}^{ikmn} X_{(a)(b)} F^{(a)}_{ik}
F_{mn}^{(b)} - {\Re}^{\,mn} Y_{(a)(b)}\hat{D}_m
\Phi^{(a)}\hat{D}_n\Phi^{(b)} \right\}\,,
\end{eqnarray}
where the so-called susceptibility tensors ${\cal R}^{ikmn}$ and
$\Re^{\,mn}$ are defined as follows:
\begin{equation}
{\cal R}^{ikmn} \equiv
\frac{q_1}{2}R\,(g^{im}g^{kn}{-}g^{in}g^{km}) {+}
\frac{q_2}{2}(R^{im}g^{kn} {-} R^{in}g^{km} {+} R^{kn}g^{im}
{-}R^{km}g^{in}) {+} q_3 R^{ikmn}\,, \label{sus}
\end{equation}
\begin{equation}\label{Re}
\Re^{\,mn}\equiv {q_4}Rg^{mn}+q_5 R^{mn}\,.
\end{equation}
Here $g = {\rm det}(g_{ik})$ is the determinant of a metric tensor
$g_{ik}$, $R$ is the Ricci scalar, $R_{mn}$ is the Ricci tensor,
$R^{i}_{\ klm}$ is the Riemann tensor. Latin indices run from 0 to
3, the group index $(a)$ runs from $1$ to $N^2-1$. $F_{mn}^{(a)}$
is the tensor of the Yang-Mills real field strength (see, e.g.,
\cite{Rubakov,Odin})
\begin{equation}
F^{(a)}_{mn} {=} \nabla_m A^{(a)}_n {-} \nabla_n A^{(a)}_m + {\cal
G} f^{(a)}_{\ (b)(c)} A^{(b)}_m A^{(c)}_n \,, \label{Fmn}
\end{equation}
where $A^{(a)}_i$ is the Yang-Mills field potential four-vector,
$\nabla_k$ is the covariant derivative. The symbol $\Phi^{(a)}$
denotes the multiplet of real scalar fields ($SU(N)$ symmetric
Higgs fields). The gauge covariant derivative $\hat{D}_m
\Phi^{(a)}$ is defined as \cite{Rubakov}
\begin{equation}
\hat{D}_m \Phi^{(a)} \equiv \nabla_m \Phi^{(a)} + {\cal G}
f^{(a)}_{\ (b)(c)} A^{(b)}_m \Phi^{(c)} \,. \label{DPhi}
\end{equation}
$m$ is a mass the Higgs field, and $q_1$, $q_2$, ... $q_5$ are the
constants of non-minimal coupling. In the definitions we follow
the book \cite{Rubakov}, i.e., we use the following basic
formulas:
\begin{equation}
G_{(a)(b)} \equiv 2 {\rm Tr} \ {\bf t}_{(a)} {\bf t}_{(b)}  \,,
\quad  \left[ {\bf t}_{(a)} , {\bf t}_{(b)} \right] = i f^{(c)}_{\
(a)(b)} {\bf t}_{(c)} \,, \label{fabc}
\end{equation}
\begin{equation}
f_{(c)(a)(b)} \equiv G_{(c)(d)} f^{(d)}_{\ (a)(b)} = - 2 i \ {\rm
Tr} \ \left[ {\bf t}_{(a)} , {\bf t}_{(b)} \right] {\bf t}_{(c)}
\,, \label{fabc1}
\end{equation}
\begin{equation}
{\bf F}_{mn} = - i {\cal G} {\bf t}_{(a)} F^{(a)}_{mn} \,, \quad
{\bf A}_m = - i {\cal G} {\bf t}_{(a)} A^{(a)}_m \,, \quad {\bf
\Phi} = {\bf t}_{(a)} \Phi^{(a)} \,. \label{represent}
\end{equation}
The symmetric tensor $G_{(a)(b)}$ plays a role of a metric in the
group space, ${\bf t}_{(a)}$ are the Hermitian traceless
generators of $SU(N)$ group, $f^{(d)}_{\ (a)(b)}$ are the
structure constants of the $SU(N)$ group, the constant ${\cal G}$
is the strength of the gauge coupling. The quantities $X_{(a)(b)}$
and $Y_{(a)(b)}$ have the following structure:
\begin{equation}
X_{(a)(b)} \equiv G_{(a)(b)} + (Q_1-1) \frac{\Phi_{(a)}
\Phi_{(b)}}{\Phi^2} \,, \quad Y_{(a)(b)} \equiv G_{(a)(b)} +
(Q_2-1) \frac{\Phi_{(a)} \Phi_{(b)}}{\Phi^2} \,, \label{XYZ}
\end{equation}
with two new coupling constants $Q_1$ and $Q_2$. When $Q_1=Q_2=1$,
the tensors $X_{(a)(b)}$ and $Y_{(a)(b)}$ coincide with the metric
of the group space. When $Q_1=0$, $X_{(a)(b)}=P_{(a)(b)}$, and
when $Q_2=0$, $Y_{(a)(b)}=P_{(a)(b)}$, where
\begin{equation}
P_{(a)(b)} \equiv G_{(a)(b)} - \frac{\Phi_{(a)}
\Phi_{(b)}}{\Phi^2} \,, \label{7para6}
\end{equation}
is a projector in the group space, which possesses the properties:
\begin{equation}
P_{(a)(b)} =  P_{(b)(a)} \,, \quad P_{(a)(b)} \Phi^{(b)} = 0 \,,
\quad P_{(a)(b)} P^{(a)(c)} = P^{(c)}_{(b)} \,, \quad
P^{(a)}_{(a)} = N^2 - 2 \,. \label{7para7}
\end{equation}
We assume here that the multiplet $\Phi^{(a)}$ has positive norm
$G_{(a)(b)}\Phi^{(a)}\Phi^{(b)} \equiv  \Phi^2
> 0$, allowing us to equip the group space by the vector $q^{(a)} =
\Phi^{(a)}/ \Phi$ with the unit norm $G_{(a)(b)} q^{(a)}
q^{(b)}=1$. Finally, the tensor $F^{ik}_{(a)}$ satisfies the
relation
\begin{equation}
\hat{D}_k F^{*ik}_{(a)} = 0 \,, \quad F^{*ik}_{(a)} =
\frac{1}{2}\epsilon^{ikls} F_{ls (a)} \,, \label{Aeq2}
\end{equation}
where $\epsilon^{ikls} = \frac{1}{\sqrt{-g}} E^{ikls}$ is the
Levi-Civita tensor, $E^{ikls}$ is the completely skew - symmetric
symbol with $E^{0123} = - E_{0123} = 1$.

\subsection{Non-minimal extension of the Yang-Mills equations}

The variation of the action functional over the Yang-Mills
potential $A^{(a)}_i$ yields
\begin{equation}
\hat{D}_k {\cal H}^{ik}_{(a)}  =   {\cal G} (\hat{D}_k
\Phi^{(d)})f^{(b)}_{\ (a)(h)} \Phi^{(h)}\left[ G_{(b)(d)} g^{ik} +
Y_{(b)(d)} \Re^{ik}  \right] \,, \label{Heqs}
\end{equation}
where the tensor ${\cal H}^{ik}_{(a)}$ is
\begin{equation}
{\cal H}^{ik}_{(a)} = F^{ik}_{(a)} + {\cal R}^{ikmn} X_{(a)(b)}
F^{(b)}_{mn} \equiv C^{ikmn}_{(a)(b)} F_{mn}^{(b)}\,. \label{HikR}
\end{equation}
Thus, the linear response tensor $C^{ikmn}_{(a)(b)}$, linking the
Yang-Mills field strength $F_{mn}^{(b)}$ and the so-called color
induction tensor ${\cal H}^{ik}_{(a)}$, is of the form
\begin{equation}
C^{ikmn}_{(a)(b)} \equiv \left[ \frac{1}{2}( g^{im} g^{kn} -
g^{in} g^{km}) + {\cal R}^{ikmn} \right] G_{(a)(b)} + (Q_1-1)
\frac{\Phi_{(a)} \Phi_{(b)}}{\Phi^2} {\cal R}^{ikmn}   \,.
\label{HikR2}
\end{equation}
The definitions of the tensors of color
permittivity, $\varepsilon^{im}_{(a)(b)}$,
color impermeability, $(\mu^{-1})^{pq}_{(a)(b)}$ and color
cross-effects, $\nu_{(a)(b)}^{p m}$, are the following
\begin{equation}
\varepsilon^{im}_{(a)(b)} {=} 2 {\cal C}^{ikmn}_{(a)(b)} U_k U_n
\,, \quad (\mu^{-1})^{pq}_{(a)(b)} {=} {-} \frac{1}{2} \eta^p_{\
ik} {\cal C}^{ikmn}_{(a)(b)} \eta^{ \ \ \ q}_{mn} \,, \quad
\nu_{(a)(b)}^{p m} {=} \eta^p_{\ ik} {\cal C}^{ikmn}_{(a)(b)} U_n
\,, \label{nu}
\end{equation}
where $\eta^{ikm} \equiv \epsilon^{ikmn} U_n$.
Using (\ref{HikR2}) and the standard definition of the projector,
$\Delta^{ik} \equiv g^{ik} - U^i U^k$, we obtain explicitly
\begin{equation}
\varepsilon^{im}_{(a)(b)} = \left[\Delta^{im}+ 2 {\cal R}^{ikmn}
U_k U_n  \right]G_{(a)(b)} + 2 (Q_1-1) {\cal R}^{ikmn} U_k U_n
\frac{\Phi_{(a)} \Phi_{(b)}}{\Phi^2}   \,, \label{Refa}
\end{equation}
\begin{equation}
(\mu^{-1})^{pq}_{(a)(b)} = \left[\Delta^{pq} - 2 \ ^{*}{\cal
R}^{*plqs} U_l U_s \right] G_{(a)(b)} - 2 (Q_1-1) \ ^{*}{\cal
R}^{*plqs} U_l U_s  \frac{\Phi_{(a)} \Phi_{(b)}}{\Phi^2} \,,
\label{Rmu}
\end{equation}
\begin{equation}
\nu^{pm}_{(a)(b)} =  - ^{*}{\cal R}^{plnm} U_l U_n
\left[G_{(a)(b)} + (Q_1-1) \frac{\Phi_{(a)} \Phi_{(b)}}{\Phi^2}
\right] \,. \label{Rnu}
\end{equation}
Assuming that there is only one preferred direction in the group
space and this direction is associated with the vector $q^{(a)}$,
one can decompose $C^{ikmn}_{(a)(b)}$ as
\begin{equation}
C^{ikmn}_{(a)(b)} = q_{(a)} q_{(b)} C^{ikmn}_{({\rm long})} +
P_{(a)(b)} C^{ikmn}_{({\rm trans})}\,, \label{ekh4}
\end{equation}
where the longitudinal and transversal parts are, respectively
\begin{equation}
C^{ikmn}_{({\rm long})} \equiv C^{ikmn}_{(a)(b)} q^{(a)} q^{(b)} =
\left[ \frac{1}{2}( g^{im} g^{kn} - g^{in} g^{km}) + Q_1 {\cal
R}^{ikmn} \right] \,, \label{lo1}
\end{equation}
\begin{equation}
C^{ikmn}_{({\rm trans})} \equiv \frac{1}{(N^2-2)}
C^{ikmn}_{(a)(b)} P^{(a)(b)}= \left[ \frac{1}{2}( g^{im} g^{kn} -
g^{in} g^{km}) + {\cal R}^{ikmn} \right] \,. \label{tr1}
\end{equation}
Mention that $C^{ikmn}_{({\rm trans})}$ can be obtained from
$C^{ikmn}_{({\rm long})}$ by the formal replacement $Q_1 \to 1$,
and let us use this features below for the simplifications of the
formulas. The corresponding longitudinal and transversal
components of $\varepsilon^{im}_{(a)(b)}$,
$(\mu^{-1})^{pq}_{(a)(b)}$ and $\nu^{pm}_{(a)(b)}$ can be easily
written in analogy with (\ref{lo1}) and (\ref{tr1}).

\subsection{Non-minimal extension of the Higgs field equations}

The variation of the action $S_{({\rm NMEYMH})}$ over the Higgs
scalar field $\Phi^{(a)}$ yields
$$
\hat{D}_m \left\{ \left[ g^{mn} G_{(a)(b)} + \Re^{mn} Y_{(a)(b)}
\right] \hat{D}_n \Phi^{(b)} \right\} =
$$
$$
- m^2 \Phi_{(a)} - \frac{(Q_1{-}1)}{2\Phi^2} \ {\cal
R}^{ikmn}F^{(c)}_{ik}F^{(b)}_{mn} \Phi_{(b)} \left[ G_{(a)(c)} -
\frac{\Phi_{(a)}\Phi_{(c)}}{\Phi^2} \right] +
$$
\begin{equation}
+ \frac{(Q_2{-}1)}{\Phi^2} \ \Re^{mn} \left[ G_{(a)(c)} -
\frac{\Phi_{(a)}\Phi_{(c)}}{\Phi^2} \right] \Phi_{(b)} (\hat{D}_m
\Phi^{(c)}) (\hat{D}_n \Phi^{(b)}) \,. \label{higgs1}
\end{equation}
This equation, clearly, has a form
\begin{equation}
\hat{D}_m \left[ {\cal C}^{mn}_{(a)(b)} \hat{D}_n \Phi^{(b)}
\right] = {\cal I}_{(a)} \,, \label{higgs12}
\end{equation}
where the tensor ${\cal C}^{mn}_{(a)(b)}$ is
\begin{equation}
{\cal C}^{mn}_{(a)(b)} = \left[g^{mn} + \Re^{mn} \right]G_{(a)(b)}
+ (Q_2-1)\Re^{mn} \frac{\Phi_{(a)} \Phi_{(b)}}{\Phi^2} \,,
\label{higgs111}
\end{equation}
and ${\cal I}_{(a)}$ stands for the right-hand-side of
(\ref{higgs1}). ${\cal C}^{mn}_{(a)(b)}$ can also be decomposed
into longitudinal and transversal components
\begin{equation}
{\cal C}^{ik}_{(a)(b)} = {\cal C}^{ik}_{({\rm long})} q_{(a)}
q_{(b)} + P_{(a)(b)}{\cal C}^{ik}_{({\rm trans})} \,, \label{hi1}
\end{equation}
where
\begin{equation}
{\cal C}^{ik}_{({\rm long})} \equiv {\cal C}^{ik}_{(a)(b)}q^{(a)}
q^{(b)} \,, \quad {\cal C}^{ik}_{({\rm trans})} \equiv
\frac{1}{(N^2-2)}{\cal C}^{ik}_{(a)(b)} P^{(a)(b)} \,.
\label{hi2}
\end{equation}
The definitions
\begin{equation}
\tilde{g}^{ik}_{({\rm long})} \equiv {\cal C}^{ik}_{({\rm long})}
= g^{ik} + Q_2 \Re^{ik} \,, \quad \tilde{g}^{ik}_{({\rm trans})}
\equiv {\cal C}^{ik}_{({\rm trans})} = g^{ik} + \Re^{ik} \,,
\label{hi4}
\end{equation}
introduce two color-acoustic metrics for the colored scalar
particles. Again, $\tilde{g}^{ik}_{({\rm trans})}$ can be
obtained from $\tilde{g}^{ik}_{({\rm long})}$ by the formal
replacement $Q_2 \to 1$.

\subsection{Master equations for the gravitational field}

The equations for the gravity field related to the action
functional $S_{({\rm NMEYMH})}$ are of the form
\begin{equation}
R_{ik} - \frac{1}{2} R \ g_{ik} = \Lambda \ g_{ik} + \kappa \left[
T^{(YM)}_{ik} + T^{(H)}_{ik} + T^{(NM)}_{ik} \right] \,.
\label{Ein}
\end{equation}
The term $T^{(YM)}_{ik}$:
\begin{equation}
T^{(YM)}_{ik} \equiv \frac{1}{4} g_{ik} F^{(a)}_{mn}F^{mn}_{(a)} -
F^{(a)}_{in}F_{k (a)}^{\ n} \,, \label{TYM}
\end{equation}
is a stress-energy tensor of the pure Yang-Mills field, the term
$T^{(H)}_{ik}$:
\begin{equation}
T^{(H)}_{ik} \equiv -  \frac{1}{2} g_{ik} \hat{D}_m \Phi^{(a)}
\hat{D}^m \Phi_{(a)} + \hat{D}_i \Phi^{(a)} \hat{D}_k \Phi_{(a)} +
\frac{1}{2} g_{ik} m^2 \Phi^{(a)} \Phi_{(a)} \, \label{TH}
\end{equation}
is a stress-energy tensor for the scalar Higgs field. The
non-minimal contributions enter the last tensor $T^{(NM)}_{ik}$,
which may be represented as a sum of five items:
\begin{equation}
T^{(NM)}_{ik} \equiv  q_1 T^{(I)}_{ik} {+} q_2 T^{(II)}_{ik} {+}
q_3 T^{(III)}_{ik} {+} q_4 T^{(IV)}_{ik} {+} q_5 T^{(V)}_{ik} \,.
\label{Tdecomp1}
\end{equation}
The definitions of these five tensors relate to the corresponding
coupling constant $q_1$,$q_2,...q_5$. The tensors $T^{(I)}_{ik}$,
$T^{(II)}_{ik}$, ... $T^{(V)}_{ik}$ are
$$
T^{(I)}_{ik} = R X_{(a)(b)}
\left[\frac{1}{4}g_{ik}F^{(a)}_{mn}F^{mn (b)} -
F^{(a)}_{im}F_{k}^{\ m (b)} \right] -  \frac{1}{2} R_{ik}
X_{(a)(b)} F^{(a)}_{mn}F^{mn (b)} +
$$
\begin{equation}
+ \frac{1}{2} \left[ \hat{D}_{i} \hat{D}_{k} - g_{ik} \hat{D}^l
\hat{D}_l \right] \left[X_{(a)(b)}F^{(a)}_{mn}F^{mn (b)}\right]
\,, \label{TI}
\end{equation}
$$
T^{(II)}_{ik} = - \frac{1}{2}g_{ik}\left[ \hat{D}_{m}
\hat{D}_{l}\left(X_{(a)(b)}F^{mn (a)}F^{l (b)}_{\ n} \right) -
R_{lm}X_{(a)(b)}F^{mn (a)}F^{l (b)}_{\ n}\right] -
$$
$$
- F^{ln (a)} X_{(a)(b)}\left(R_{il}F^{(b)}_{kn} +
R_{kl}F^{(b)}_{in}\right) - \frac{1}{2} \hat{D}^m \hat{D}_m
\left(X_{(a)(b)}F^{(a)}_{in}F_{k}^{\ n (b)}\right) +
$$
\begin{equation}
+  \frac{1}{2}\hat{D}_l \left[ \hat{D}_i \left(
X_{(a)(b)}F^{(a)}_{kn}F^{ln (b)}\right) + \hat{D}_k
\left(X_{(a)(b)}F^{(a)}_{in}F^{ln (b)}\right) \right]  -
R^{mn}X_{(a)(b)}F^{(a)}_{im}F^{(b)}_{kn} \,, \label{TII}
\end{equation}
$$
T^{(III)}_{ik} = \frac{1}{4}g_{ik}
R^{mnls}X_{(a)(b)}F^{(a)}_{mn}F^{(b)}_{ls} - \frac{3}{4}
X_{(a)(b)}F^{ls (a)}\left(F_{i}^{\ n (b)} R_{knls} + F_{k}^{\ n
(b)}R_{inls}\right) -
$$
\begin{equation}
- \frac{1}{2}\hat{D}_{m} \hat{D}_{n}
\left[X_{(a)(b)}\left(F_{i}^{\ n (a)}F_{k}^{\ m (b)} + F_{k}^{\ n
(a)}F_{i}^{\ m (b)}\right) \right] \,, \label{TIII}
\end{equation}
$$
T^{(IV)}_{ik} = \left(R_{ik} - \frac{1}{2}g_{ik} R \right)
Y_{(a)(b)} (\hat{D}_m \Phi^{(a)})(\hat{D}^m \Phi^{(b)}) + R
Y_{(a)(b)} (\hat{D}_i \Phi^{(a)})(\hat{D}_k \Phi^{(b)}) +
$$
\begin{equation}
+ \left( g_{ik} \hat{D}^n \hat{D}_n - \hat{D}_i \hat{D}_k \right)
\left[Y_{(a)(b)} (\hat{D}_m \Phi^{(a)})(\hat{D}^m \Phi^{(b)})
\right] \,, \label{TIV}
\end{equation}
$$
T^{(V)}_{ik} {=}  Y_{(a)(b)} (\hat{D}_m \Phi^{(b)}) \left[ R^m_i
(\hat{D}_k \Phi^{(a)}) {+} R^m_k (\hat{D}_i \Phi^{(a)}) \right]
{-} \frac{1}{2}R_{ik} Y_{(a)(b)} (\hat{D}_m \Phi^{(a)})(\hat{D}^m
\Phi^{(b)}) {-}
$$
$$
- \frac{1}{2} \hat{D}^m \left\{  \hat{D}_i \left[Y_{(a)(b)}
(\hat{D}_m \Phi^{(a)})(\hat{D}_k \Phi^{(b)}) \right] + \hat{D}_k
\left[ Y_{(a)(b)} (\hat{D}_m \Phi^{(a)})(\hat{D}_i \Phi^{(b)})
\right] - \right.
$$
\begin{equation}
\left.  - \hat{D}_m \left[ Y_{(a)(b)} (\hat{D}_i
\Phi^{(a)})(\hat{D}_k \Phi^{(b)}) \right] \right\} +
\frac{1}{2}g_{ik} \hat{D}_m \hat{D}_n \left[Y_{(a)(b)}
\left(\hat{D}^m \Phi^{(a)}\right) \left(\hat{D}^n
\Phi^{(b)}\right) \right]
 \,. \label{TV}
\end{equation}
Presented equations generalize the gravity field equations in the
three-parameter model considered in \cite{1BZ06,2BZ06,BSZ07}.

\section{Application of the effective metric formalism}

\subsection{pp-wave background}

Consider now one specific plane-symmetric spacetime associated
usually with a gravitational radiation. We assume the metric to be
of the form
\begin{equation}
ds^2 = 2dudv -L^2(u) \left[e^{2\beta(u)} (dx^2)^2 + e^{-2\beta(u)}
(dx^3)^2 \right]
 \,, \label{pp1}
\end{equation}
where $u= (t-x^1)/\sqrt2$ and $v= (t+x^1)/\sqrt2$ are the retarded
and advanced time, respectively. This spacetime is known to admit
the $G_5$ group of isometries \cite{ExactSolution}, and three
Killing four-vectors, $\xi^k$, $\xi^k_{(2)}$ and $\xi^k_{(3)}$
form three-dimensional Abelian subgroup $G_3$. The four-vector
$\xi^k$ is the null one and covariantly  constant, i.e.,
\begin{equation}
\xi^k = \delta^k_{v} \,, \quad g_{kl} \ \xi^k \xi^l = 0 \,, \quad
\nabla_l \ \xi^k =0 \,. \label{pp2}
\end{equation}
The four-vectors $\xi^k_{(\alpha)}$ (here and below $\alpha = 2,
3$) are space-like and orthogonal to $\xi^k$ and each others,
i.e.,
\begin{equation}
\xi^k_{(\alpha)} = \delta^k_{\alpha} \,, \quad g_{kl} \
\xi^k_{(2)} \xi^l_{(3)} = 0 \,, \quad g_{kl} \ \xi^k
\xi^l_{(\alpha)} = 0 \,. \label{pp3}
\end{equation}
Two tetrad vectors, $X^k_{(\alpha)}$, are associated with
$\xi^k_{(\alpha)}$
\begin{equation}
X^k_{(2)} = \delta^k_{2} \  \frac{e^{-\beta}}{L} \,, \quad
X^k_{(3)} = \delta^k_{3} \ \frac{e^{\beta}}{L} \,, \quad
g_{kl}X^k_{(\alpha)} \ X^l_{(\beta)} = - \delta_{(\alpha)(\beta)}
\,. \label{pp4}
\end{equation}
The non-vanishing components of the Ricci and Riemann tensors are,
respectively
\begin{equation}
R_{uu}=R^2_{ \ u2u}+R^3_{ \ u3u} \,, \quad R^2_{ \ u2u} = -
\left[\frac{L^{\prime \prime}}{L} + (\beta^{\prime})^2 \right] -
\left[2\beta^{\prime} \frac{L^{\prime}}{L} +  \beta^{\prime
\prime} \right] \,, \label{ppRie1}
\end{equation}
\begin{equation}
R^3_{ \ u3u} = - \left[\frac{L^{\prime \prime}}{L} +
(\beta^{\prime})^2 \right] + \left[2\beta^{\prime}
\frac{L^{\prime}}{L} +  \beta^{\prime \prime} \right]\,.
\label{ppRie2}
\end{equation}
We consider a {\it toy-model}, which satisfies the following
requirements. {\it First}, the background Yang-Mills and Higgs
fields are parallel in the group space \cite{Yasskin,Galtsov},
i.e.,
\begin{equation}
A^{(a)}_k = q^{(a)} A_k \,, \quad \Phi^{(a)} = q^{(a)} \Phi \,,
\quad G_{(a)(b)} q^{(a)}q^{(b)} = 1 \,, \quad q^{(a)} = const \,.
\label{pp5}
\end{equation}
{\it Second}, the vector field $A_k$ and scalar field $\Phi$
inherit the symmetry of the spacetime, i.e., the Lie derivatives
of these quantities along generators of the group $G_3$, $\{\xi \}
\equiv \{ \xi^k, \xi^k_{(2)}, \xi^k_{(3)} \}$, are equal to zero:
\begin{equation}
\pounds_{\{\xi\}} A_k = 0 \,, \quad \pounds_{\{\xi\}} \Phi = 0 \,.
\label{pp6}
\end{equation}
{\it Third}, the vector field $A^k$ is transverse, i.e.,
\begin{equation}
A^k = - \left[ A_{(2)} X^k_{(2)} + A_{(3)} X^k_{(3)} \right]  \,,
\quad \xi^k A_k =0 \,. \label{pp7}
\end{equation}
{\it Fourth}, the background scalar field is massless. {\it Fifth},
the cosmological constant is absent, $\Lambda=0$. These five
requirements lead to the following simplifications.

\noindent {\it (i)} The fields $A_k$ and $\Phi$ depend on the
retarded time $u$ only; there are two non-vanishing components of
the field strength tensor
\begin{equation}
F^{(a)ik} = q^{(a)} \left[ \left( \xi^i \xi^k_{(2)} - \xi^k
\xi^i_{(2)} \right) A^{\prime}_{2}(u) + \left(\xi^i \xi^k_{(3)} -
\xi^k \xi^i_{(3)}\right) A^{\prime}_{3}(u) \right] \,,
\end{equation}
the invariant $F^{(a)}_{ik}F_{(a)}^{ik}$ as well as the terms
${\cal R}^{ikmn}F_{mn}^{(a)}$ are equal to zero; there is only one
non-vanishing component of the derivative $\nabla_k \Phi$, namely,
$\partial_u \Phi$.

\noindent {\it (ii)} The expression in the right-hand-side of
(\ref{Heqs}) vanishes, ${\cal H}^{ik}_{(a)}$ coincides with
$F^{ik}_{(a)}$; the equations (\ref{Heqs}) and (\ref{Aeq2}) are
satisfied identically.

\noindent {\it (iii)} The right-hand-side of the equation
(\ref{higgs1}) vanishes and this equation is satisfied
identically.

\noindent {\it (iv)} All the non-minimal terms $T^{(I)}_{ik}$,
..., $T^{(V)}_{ik}$ (\ref{TI})-(\ref{TV}) disappear, and the
coupling parameters $q_1$, $q_2$,..., $q_5$, being non-vanishing,
happen to be hidden, i.e., they do not enter the equations for the
gravity field.

\noindent After such simplifications the total effective
stress-energy tensor in the right-hand-side of (\ref{Ein}) is the
null one (as it should be, \cite{ExactSolution}), i.e., this
tensor has a structure $\xi_i \xi_k T(u)$. Thus, the non-minimal
equations for the gravity field (\ref{Ein})-(\ref{TV}) reduce to
one equation
\begin{equation}
\frac{L^{\prime \prime}}{L} + (\beta^{\prime})^2 =- \frac{1}{2}
\kappa T(u)\,, \label{pp8}
\end{equation}
where
\begin{equation}
T(u) = \frac{1}{L^2} \left[e^{-2\beta(u)}
\left(A^{\prime}_{2}\right)^2(u) + e^{2\beta(u)}
\left(A^{\prime}_{3}\right)^2(u)\right] +
\left(\Phi^{\prime}(u)\right)^2 \geq 0 \,, \label{pp9}
\end{equation}
$A_2$, $A_3$, $\Phi$ are arbitrary functions of the retarded time
$u$, and the prime denotes the derivative with respect to retarded
time. As usual, we can treat $\beta(u)$ as arbitrary function and
find $L(u)$ as a solution of (\ref{pp8}).

\subsection{Structure of the linear response tensor}

Taking into account the formula (\ref{ekh4}) introducing two
material tensors: the longitudinal one, $C^{ikmn}_{({\rm long})}$,
and the transversal one, $C^{ikmn}_{({\rm trans})}$, we have to
focus now on their decomposition for the case of pp-wave
background. Mention that if we have a representation of the
transversal linear response tensor, related to the coupling
constants $q_1$, $q_2$ and $q_3$, the representation of the
longitudinal one can be obtained by simple replacement $q_1 \to
Q_1 q_1$, $q_2 \to Q_1 q_2$ and $q_3 \to Q_1 q_3$. Thus, below we
consider the formulas for $C^{ikmn}_{({\rm trans})}$ only. They
are
\begin{equation}
C^{ikmn}_{({\rm trans})} = \frac{1}{2}\left(g^{im} g^{kn} -
g^{in}g^{km}\right) + \left(\Pi^{im} \xi^{k}\xi^{n}  -
\Pi^{in}\xi^{k}\xi^{m} + \Pi^{kn}\xi^{i}\xi^{m} -
\Pi^{km}\xi^{i}\xi^{n} \right) \,, \label{pp10}
\end{equation}
where
\begin{equation}
\Pi^{im} \equiv \frac{1}{2} q_2 \left(R^2_{\ u2u} + R^3_{\ u3u}
\right) g^{im} - q_3 \left[R^2_{\ u2u} X^{i}_{(2)} X^{m}_{(2)} +
R^3_{\ u3u} X^{i}_{(3)}X^{m}_{(3)} \right] \,. \label{pp11}
\end{equation}
Clearly, the velocity four-vector
$U^k=\frac{1}{\sqrt{2}}(\delta^k_u+\delta^k_v)$ is an eigenvector
of the tensor $\Pi^{im}$, and $\Pi$ is the corresponding
eigenvalue:
\begin{equation}
\Pi^{im}U_m = \Pi U^i \,, \quad \Pi \equiv \frac{1}{2} q_2
\left(R^2_{\ u2u} + R^3_{\ u3u} \right) = \frac{1}{2} \kappa q_2 T
\,, \label{pp13}
\end{equation}
thus, the permittivity tensor, impermeability tensor and
cross-effect tensor are, respectively, (we omit here the mark
``trans" for simplicity)
\begin{equation}
\varepsilon^{im} = \Delta^{im} + \Pi^{im} + 2 \Pi \xi^i \xi^m -
\sqrt2 \Pi (U^i \xi^m + U^m \xi^i) \,, \label{pp12}
\end{equation}
\begin{equation}
\left(\mu^{-1}\right)^{pq} = \Delta^{pq} - 2 \eta^p_{ \ ik}
\eta^q_{ \ mn} \Pi^{im} \xi^k \xi^n \,, \quad \nu^{pm} = \sqrt2 \
\eta^p_{\ ik} \ \xi^k \ \Pi^{im} \,. \label{Rnu1}
\end{equation}
When the observer is at rest, and its velocity four-vector is $U^k
= \delta^k_0 = (\delta^k_u + \delta^k_v )/ \sqrt2$, the
non-vanishing permittivity, impermeability and magneto-electric
coefficients can be easily calculated:
\begin{equation}
\varepsilon^{2}_{2} = 1 + \Pi^2_2 \,, \quad \varepsilon^{3}_{3} =
1 + \Pi^3_3 \,, \quad (\mu^{-1})^{2}_{2} = 1 - \Pi^3_3 \,, \quad
(\mu^{-1})^{3}_{3} = 1 - \Pi^2_2 \quad  \,, \label{pp15}
\end{equation}
\begin{equation}
\nu^{23} = \frac{1}{L^2} \Pi^3_3 \,, \quad \nu^{32} = -
\frac{1}{L^2} \Pi^2_2 \,, \label{Rnu2}
\end{equation}
where
\begin{equation}
\Pi^2_2 \equiv \Pi + q_3 R^2_{\ u2u} \,, \quad \Pi^3_3 \equiv \Pi
+ q_3 R^3_{\ u3u} \,. \label{pp14}
\end{equation}
There are two principal cases in this model.

\vspace{3mm}

\noindent {\it (i) } $\Pi^2_2 \neq \Pi^3_3$

\noindent For such case $\varepsilon^{2}_{2} \neq
\varepsilon^{3}_{3}$ and $(\mu^{-1})^{2}_{2} \neq
(\mu^{-1})^{3}_{3}$, thus the symmetry of the model can not be
considered as uniaxial. The interesting particular case here is
$q_2+q_3=0$, which relates to $\Pi^2_2 = - \Pi^3_3$. When we
reconstruct the effective metrics for the pp-wave model of the
non-minimally active vacuum, we can use neither the results
obtained by Perlick \cite{Perlick} (since the cross-effects of
magneto-electricity are present), nor the explicit results of
\cite{GRG05} (since the model is not uniaxial). Below we solve
this problem using an alternative way.

\vspace{3mm}

\noindent {\it (ii) } $\Pi^2_2 = \Pi^3_3$

\noindent This special case can be treated as uniaxial with
antisymmetric tensor of cross effects. It is possible when
\begin{equation}
R^{2}_{ \ u2u} = R^{3}_{ \ u3u} \ \ \rightarrow 2 \beta^{\prime}
\frac{L^{\prime}}{L} + \beta^{\prime \prime} = 0  \ \ \rightarrow
\beta^{\prime} = \frac{const}{L^2} \,. \label{ukh}
\end{equation}
In addition, if $q_2+q_3=0$, one obtains $\Pi^2_2 = \Pi^3_3 = 0$,
and the non-minimal interactions do not influence the wave
propagation in the vacuum.

\subsection{Reconstruction of the effective metrics}

Since the spacetime remains active from the point of view of
curvature coupling (even if coupling parameters are hidden for
the background fields), we can now construct the effective metrics
for the {\it test particles}, treating the  non-minimal
interactions as a consequence of presence of some effective
medium with planar spatial symmetry. This means that we take,
first, the material tensor $C^{ikmn}_{(a)(b)}$ and decompose it
algebraically with respect to associated metrics. Then we
consider the Yang-Mills and Higgs equations in the WKB
approximation and obtain color and color-acoustic metrics,
respectively.

\subsubsection{Associated metrics}

Let us introduce three symmetric linearly independent tensor
fields
$$
h^{im}_{(I)} = g^{im} + 2 \Pi \xi^i \xi^m \,, \quad h^{im}_{(II)}
= \xi^i \xi^m + 2 (\Pi - \Pi^2_2) X^i_{(2)}X^m_{(2)} \,,
$$
\begin{equation}
h^{im}_{(III)} = \xi^i \xi^m  + 2 (\Pi - \Pi^3_3)
X^i_{(3)}X^m_{(3)} \,. \label{M1}
\end{equation}
Using these quantities one can show directly, that
\begin{equation}
2 C^{ikmn}_{({\rm trans})} {=} \left[ h^{im}_{(I)} h^{kn}_{(I)}
{-} h^{in}_{(I)} h^{km}_{(I)} \right] {+} \left[ h^{im}_{(II)}
h^{kn}_{(II)} {-} h^{in}_{(II)} h^{km}_{(II)} \right] {+} \left[
h^{im}_{(III)} h^{kn}_{(III)} {-} h^{in}_{(III)} h^{km}_{(III)}
\right] \,. \label{M2}
\end{equation}
This formula is a particular case of the multi-metric
representation \cite{GRG05}
\begin{equation}
C^{ikmn} = \frac{1}{2\hat{\mu}}\sum_{(\alpha)(\beta)}
G_{(\alpha)(\beta)} \left[g^{im (\alpha)} \ g^{kn (\beta)} - g^{in
(\alpha)} \ g^{km (\beta)} \right] \label{supergeneral}
\end{equation}
with $\hat{\mu}=1$, $G_{(\alpha)(\beta)} =
\delta_{(\alpha)(\beta)}$ and $(\alpha), (\beta) = (I),(II),(II)$.
Thus, the tensor fields $h^{im}_{(I)}$, $h^{im}_{(II)}$ and
$h^{im}_{(III)}$ can be indicated as metrics {\it associated} with
$C^{ikmn}_{({\rm trans})}$. The metrics, associated with
$C^{ikmn}_{({\rm long})}$, say, $\tilde{h}^{im}_{(I)}$,
$\tilde{h}^{im}_{(II)}$ and $\tilde{h}^{im}_{(III)}$, can be
defined analogously with the replacement $q_2 \to Q_1 q_2$ and
$q_3 \to Q_1 q_3$. The attempts to decompose $C^{ikmn}_{({\rm
long})}$ and $C^{ikmn}_{({\rm trans})}$ by using {\it two}
associated metrics face with mathematical contradiction. Thus, we
conclude, that an appropriate internal (associated) sub-space is
three-dimensional in the presented case, when vacuum interacting
with curvature possesses color cross-effects.

\subsubsection{Color metrics}

In the WKB-approximation the gauge potentials $A_k^{(a)}$ and the
field strength $F_{kl}^{(a)}$ can be extrapolated as follows
\begin{equation}
A_k^{(a)} \rightarrow {\cal A}_k^{(a)} e^{i \Psi} \,, \quad
F_{kl}^{(a)} \rightarrow i \left[ p_k {\cal A}_l^{(a)} - p_l {\cal
A}_k^{(a)}\right] e^{i \Psi} \,, \quad p_k = \nabla_k \Psi
\,.\label{AP1}
\end{equation}
Mention that the nonlinear terms in (\ref{Fmn}) give the values of
the next order in WKB approximation, thus, such a model of gauge
field is effectively Abelian. In the leading order approximation
the Yang-Mills equations reduce to
\begin{equation}
C^{ikmn}_{(a)(b)} \ p_k \ p_m \ {\cal A}_n^{(b)} = 0 \,.
\label{AP2}
\end{equation}
Projection (\ref{AP2}) onto direction $q^{(a)}$ yields
\begin{equation}
C^{ikmn}_{({\rm long})} \ p_k \ p_m \ {\cal A}_n^{(||)} = 0 \,,
\quad {\cal A}_n^{(||)} \equiv {\cal A}_n^{(b)} q_{(b)} \,.
\label{AP211}
\end{equation}
Convolution of the equations (\ref{AP2}) with $P^{(a)(c)}$ gives
\begin{equation}
C^{ikmn}_{({\rm trans})} \ p_k \ p_m \ {\cal A}_n^{(c)(\bot)} = 0
\,, \quad {\cal A}_n^{(c)(\bot)} \equiv {\cal A}_n^{(b)}
P^{(c)}_{(b)} \,. \label{AP212}
\end{equation}
Propagation of longitudinal (with respect to direction pointed by
$q^{(a)}$) and transversal color waves, presented by the
quantities ${\cal A}_n^{(||)}$ and ${\cal A}_n^{(c)(\bot)}$,
respectively, can be described by the same method, but the
analysis is based on the decomposition of $C^{ikmn}_{({\rm
long})}$ or $C^{ikmn}_{({\rm trans})}$, correspondingly. Below we
analyse the {\it transversal} case only, the longitudinal one can
be described using the replacement $q_2 \to Q_1 q_2$ and $q_3 \to
Q_1 q_3$.

Let us project the equation (\ref{AP212}) onto the direction given
by $\xi^k$. This procedure yields the scalar relation
\begin{equation}
\left(\xi^k p_k \right) \left[p^l {\cal A}_l^{(c)(\bot)} \right] -
\left(p^k p_k \right)\left[\xi^l {\cal A}_l^{{(c)(\bot)}} \right]
= 0 \,. \label{AP3}
\end{equation}
As usual, we consider the gauge condition of the Landau-type
$\xi^l {\cal A}_l^{{(c)(\bot)}} = 0$ for the Yang-Mills potential
four-vector, and obtain immediately, that $p^l {\cal
A}_l^{{(c)(\bot)}}=0$, i.e., the particle four-momentum is
orthogonal to the four-vector of the field amplitude. Then we
project the equation (\ref{AP212}) onto the directions given by
$X^k_{(2)}$ and $X^k_{(3)}$ and obtain, respectively,
\begin{equation}
{\cal A}_{(2)}^{{(c)(\bot)}}  \left\{ g^{ik} + \xi^i \xi^k \left[
q_2 \left(R^2_{\ u2u} + R^3_{\ u3u} \right) + 2q_3 R^2_{\ u2u}
\right] \right\} p_i p_k =0\,, \label{AP4}
\end{equation}
\begin{equation}
{\cal A}_{(3)}^{{(c)(\bot)}} \left\{ g^{ik} + \xi^i \xi^k \left[
q_2 \left(R^2_{\ u2u} + R^3_{\ u3u} \right) + 2q_3 R^3_{\ u3u}
\right] \right\} p_i p_k =0\,, \label{AP5}
\end{equation}
where ${\cal A}_{(2)}^{{(c)(\bot)}} \equiv X^m_{(2)}{\cal
A}_{m}^{{(c)(\bot)}}$, etc. Clearly, the color metrics for the
waves with the polarization ${\cal A}_{(2)}^{{(c)(\bot)}}\neq 0$,
${\cal A}_{(3)}^{{(c)(\bot)}}= 0$, and vice versa are,
respectively
\begin{equation} \label{AP7}
g^{ik}_{(2)} = g^{ik} + 2 \xi^i \xi^k \Pi^2_2 = g^{ik} +  \xi^i
\xi^k \left[ \varepsilon^2_2 - \frac{1}{\mu^3_3} \right] \,,
\end{equation}
\begin{equation} \label{AP8}
g^{ik}_{(3)} = g^{ik} + 2 \xi^i \xi^k \Pi^3_3 = g^{ik} +  \xi^i
\xi^k \left[ \varepsilon^3_3 - \frac{1}{\mu^2_2}  \right] \,.
\end{equation}
When $R^2_{\ u2u} \neq - R^3_{\ u3u}$, i.e., $R_{uu} \neq 0$ and
$T(u)\neq 0$, these color metrics can be expressed as a linear
combinations of $g^{ik}$ and $h^{ik}_{(I)}$. When the wave with
mixed polarization propagates in the described background, it
splits into two waves, moving with different phase velocities.
This is the effect of the gravitationally induced birefringence,
described, first, by Drummond and Hathrell \cite{Drum} for the
case of weak pure gravitational wave, and then investigated in
detail for the non-linear case in
\cite{Balakin1,B1,B2,B4,Balakin4}. The results given by
(\ref{AP7}) and (\ref{AP8}) are generalizations of that ones for
the case, when the background spacetime is not empty. In case
with empty background spacetime the unique component of the Ricci
tensor $R_{uu} = R^2_{\ u2u} + R^3_{\ u3u}$ and, consequently,
the quantity $\Pi$, were equal to zero,  and the equality
$\Pi^2_2 = - \Pi^3_3$ took place. Thus, when the spacetime is
empty and corresponds to the so-called pure pp-wave gravitational
background, one of the waves, say, wave with the polarization
along $Ox^2$, is subluminal, but the second one, with the
orthogonal polarization, is superluminal (i.e., the wave phase
velocities are more or less than speed of light in pure vacuum).
In our case the situation is more sophisticated. Depending on the
signs and values of the coupling parameters $q_2$ and $q_3$ one
can obtain two additional cases: first, both waves are
subluminal, second, both wave are superluminal. When $q_2=-q_3$,
clearly, the first case is realized, i.e., one of the waves is
subluminal, the other is superluminal.

Given interpretation can be motivated by two ways. The first one
is direct and uses the dispersion relations. One can rewrite the
dispersion relations following from the equations
(\ref{AP4}),(\ref{AP5}) in terms of frequency $\omega \equiv U^k
p_k$ and components of the wave three-vector $X^k_{(a)}p_{k}$ as
follows
\begin{equation} \label{okh1}
\omega^2_{(2)} = p^2 - 2 p^2_v \Pi^2_2 \,, \quad \omega^2_{(3)} =
p^2 - 2 p^2_v \Pi^3_3 \,,
\end{equation}
where $p^2 \equiv p_1^2 - g^{\alpha \beta} p_{\alpha} p_{\beta}$
is a square of the momentum three-vector. Keeping in mind, that
the quantities $\omega_{(2)} / p$ and $\omega_{(3)} / p$ are phase
velocities of the waves with polarization along $Ox^2$ and $Ox^3$,
respectively, we can conclude that positive $\Pi^2_2$ ( or
$\Pi^3_3$) characterizes the wave with phase velocity less than
one, negative one relates to the superluminal wave. The second
interpretation is based on the analysis of the effective line
elements
\begin{equation} \label{okh2}
ds^2_{(\alpha)} \equiv g_{ik (\alpha)} dx^i dx^k \,.
\end{equation}
Clearly, the metrics on the plane $x^2Ox^3$ coincide, i.e.,
\begin{equation} \label{okh21}
ds^2_{\bot} \equiv g_{22} (dx^2)^2 + g_{33} (dx^3)^2 =
ds^2_{(2)\bot} = ds^2_{(3)\bot} \,,
\end{equation}
as for the metrics on the cross-section $tOx^1$
$$
ds^2_{||(2)} = 2 dudv - 2du^2 \ \Pi^2_2 = dt^2(1-\Pi^2_2) -
(dx^1)^2 (1+\Pi^2_2) + 2 \Pi^2_2 \ dt dx^1 \,,
$$
$$
ds^2_{||(3)} = 2 dudv - 2du^2 \ \Pi^3_3 = dt^2(1-\Pi^3_3) -
(dx^1)^2 (1+\Pi^3_3) + 2 \Pi^3_3 \ dt dx^1  \,,
$$
\begin{equation} \label{okh3}
ds^2_{||} = 2 dudv = dt^2 - (dx^1)^2 \,,
\end{equation}
they differ one from another. Let us repeat, that the analysis of
the problem for the longitudinal wave characterized by ${\cal
A}_n^{(||)}$ can be done using the same method and the results can
be obtained from described above by the replacement $q_2 \to Q_1
q_2$ and $q_3 \to Q_1 q_3$.

\subsubsection{Color-acoustic metric}

In the WKB approximation the equations for color massless scalar
fields reduce to
\begin{equation} \label{Acu01}
\left\{\tilde{g}^{ik}_{({\rm long})} q_{(a)} \left[q_{(b)}
\Phi^{(b)}\right] + \tilde{g}^{ik}_{({\rm trans})}
\left[P_{(a)(b)} \Phi^{(b)}\right] \right\} p_i p_k = 0 \,.
\end{equation}
Propagation of the longitudinal scalar field $\Phi^{(||)} \equiv
q_{(b)} \Phi^{(b)}$ is thus described by the color-acoustic metric
$\tilde{g}^{ik}_{({\rm long})}$, as for transversal scalar fields
$\Phi^{(\bot)}_{(a)} \equiv P_{(a)(b)} \Phi^{(b)}$, their
propagation is characterized by the metric $\tilde{g}^{ik}_{({\rm
trans})}$. For the pp-wave background the transversal
color-acoustic metric $\tilde{g}^{ik}_{({\rm trans})} \equiv
g^{ik} + \Re^{ik}$ takes the following form
\begin{equation} \label{Acu1}
\tilde{g}^{ik} = g^{ik} + q_5 \xi^i \xi^k \left(R^2_{ \ u2u} +
R^3_{ \ u3u} \right) = g^{ik} + \kappa q_5 \xi^i \xi^k T(u) \,.
\end{equation}
As in the case of color metrics, the true metric in the plane
$x^2Ox^3$ coincides with the color-acoustic one. Taking into
account that
\begin{equation} \label{okh4}
\tilde{ds}^2_{||} = 2 dudv - q_5 \kappa T du^2 = dt^2
\left(1-\frac{1}{2} q_5 \kappa T \right) - (dx^1)^2
\left(1+\frac{1}{2} q_5 \kappa T \right) + q_5 \kappa T \ dt dx^1
\,,
\end{equation}
we conclude again that the difference appears in the cross-section
$tOx^1$ only. The longitudinal color-acoustic metric
$\tilde{g}^{ik}_{({\rm long})} \equiv g^{ik} + Q_2\Re^{ik}$ can be
obtained from the transversal one by the replacement $q_4 \to Q_2
q_4$ and $q_5 \to Q_2 q_5$, and we omit here the corresponding
analysis.

\section{Exact solution to the non-minimal Yang - Mills equations
in the model of parallel fields}

Let us omit now the requirements of the WKB-approximation and
consider the Yang-Mills equations in the framework of the model
with parallel fields. We can use the obtained color metrics as a
hint in the searching for exact solutions to the Yang-Mills
equations in the Abelian-type model. Using the multiplicative
representation $A^{(a)}_k = q^{(a)} A_k$, we omit below the group
index $(a)$. The master equations in this case are linear
\begin{equation}
\nabla_k \left[ F^{ik} + {\cal R}^{ikmn}F_{mn} \right] = 0 \,.
\label{Ex1}
\end{equation}
The first hint is to use the Landau-type (algebraic) gauge
condition $\xi^k A_k = 0$, i.e., $A_v=0$. Then the  standard
Lorentz (differential) gauge condition gives
\begin{equation}
\nabla_k A^{k} = 0 \ \ \rightarrow \ \ \partial_v A_u + g^{\alpha
\beta} \partial_{\alpha} A_{\beta} = 0 \,, \label{Ex2}
\end{equation}
where again $\alpha, \beta = 2,3$. The equations for the
components $A_2$ and $A_3$ take the following form
\begin{equation}
 \left[ 2 \partial_{u} \partial_{v} + g^{\alpha \beta}\partial_{\alpha}\partial_{\beta}
 + 2 \Pi^2_2 \partial^2_{v} - 2 \beta^{\prime} \partial_{v} \right] A_2 = 0 \,, \label{Ex3}
\end{equation}
\begin{equation}
 \left[ 2 \partial_{u} \partial_{v} + g^{\alpha \beta}\partial_{\alpha}\partial_{\beta}
 + 2 \Pi^3_3 \partial^2_{v} + 2 \beta^{\prime} \partial_{v} \right] A_3 = 0 \,. \label{Ex4}
\end{equation}
The exact solutions of these equations are
\begin{equation}
A_2 = e^{\beta} B_2(W_{(2)}) \,, \quad  A_3 = e^{-\beta}
B_3(W_{(3)}) \,, \label{Ex5}
\end{equation}
\begin{equation}
A_u = - \frac{1}{L^2 k_v} \left[ e^{-\beta} k_2 B_2(W_{(2)}) +
e^{\beta} k_3 B_3(W_{(3)}) \right] \,,\label{Ex55}
\end{equation}
where the scalar functions $W_{(2)}$ and $W_{(3)}$, the arguments
of arbitrary functions $B_2$ and $B_3$, can be treated as phases
of the corresponding waves:
\begin{equation}
W_{(2)} = W_{(2)}(0) - \frac{k_{\alpha}k_{\beta}}{2k_v} \int_0^u
du' g^{\alpha \beta}(u') - k_v \int_0^u du' \Pi^2_2(u') +  k_v \ v
+ k_{\alpha} x^{\alpha} \,, \label{Ex6}
\end{equation}
\begin{equation}
W_{(3)} = W_{(3)}(0) - \frac{k_{\alpha}k_{\beta}}{2k_v} \int_0^u
du' g^{\alpha \beta}(u') - k_v \int_0^u du' \Pi^3_3(u') + k_v \ v
+ k_{\alpha} x^{\alpha} \,. \label{Ex7}
\end{equation}
Here the constants $k_v$, $k_2$ and $k_3$ are initial values of
the wave vector at $u=0$, when the metric (\ref{pp1}) coincides
with the Minkowski one. The wave four-vectors $K^i_{(2)}$ and
$K^i_{(3)}$, defined as
\begin{equation}
K^i_{(2)} \equiv \nabla^i W_{(2)} = - \delta^i_u \left[
\frac{k_{\alpha}k_{\beta}}{2k_v} g^{\alpha \beta} + k_v \Pi^2_2
\right] + \delta^i_v k_v + \delta^i_2 k_2 + \delta^i_3 k_3 \,,
\label{Ex8}
\end{equation}
\begin{equation}
K^i_{(3)} \equiv \nabla^i W_{(3)} = - \delta^i_u \left[
\frac{k_{\alpha}k_{\beta}}{2k_v} g^{\alpha \beta} + k_v \Pi^3_3
\right] + \delta^i_v k_v + \delta^i_2 k_2 + \delta^i_3 k_3 \,,
\label{Ex9}
\end{equation}
satisfy the relations
\begin{equation}
g_{im} K^i_{(2)} K^m_{(2)} =  - k_v^2 \Pi^2_2 \,, \quad g_{im}
K^i_{(3)} K^m_{(3)} =  - k_v^2 \Pi^3_3 \,,   \label{Ex10}
\end{equation}
\begin{equation}
g_{im} K^i_{(2)} K^m_{(3)} = - k_v^2 \left(\Pi^2_2 + \Pi^3_3
\right)\,. \label{Ex101}
\end{equation}
Clearly, the equations (\ref{Ex10}) relate to (\ref{AP4}),
(\ref{AP5}), when $K^i_{(2)}$ and $K^i_{(3)}$ are identified with
$p^i$. In this sense the color metrics $g^{ik}_{(2)}$ and
$g^{ik}_{(3)}$ in (\ref{AP7}) and (\ref{AP8}) were the hints for
the reconstruction of the phase scalars (\ref{Ex6}) and
(\ref{Ex7}), respectively. As well, the potential four-vector is
orthogonal to both $K^i_{(2)}$ and $K^i_{(3)}$:
\begin{equation}
g_{im} K^i_{(2)} A^m  = g_{im} K^i_{(3)} A^m = 0 \,. \label{Ex11}
\end{equation}
This exact solution is a generalization of the results
\cite{Balakin1} for the case, when the pp-wave spacetime is not
empty, and the parallel null Yang-Mills and Higgs fields form the
spacetime  background.

\section{Exact solution of the Higgs equation \\ in the model of parallel fields}

Consider now the equation
\begin{equation}
\nabla_k \left[\left(g^{kn} + \Re^{kn} \right)\nabla_n \Phi
\right] = - m^2 \Phi \,, \label{Ex12}
\end{equation}
for the given spacetime background. It can be rewritten as
\begin{equation}
\left[ 2 \partial_{u} \partial_{v} + g^{\alpha
\beta}\partial_{\alpha}\partial_{\beta}
 + q_5 \kappa T(u) \partial^2_{v} + 2 \frac{L^{\prime}}{L} \partial_{v} + m^2 \right] \Phi =
 0\,,
 \label{Ex13}
\end{equation}
and the exact solution can be obtained in analogy with the one for
the Yang-Mills equations. Indeed, when the mass of the test scalar
particle, $m$, is equal to zero, $m=0$, the basic solution is
\begin{equation}
\Phi {=} \frac{1}{L} B(W) \,, \quad  W {=} W(0) {+}
\frac{m^2}{2k_v} u {-} \frac{k_{\alpha}k_{\beta}}{2k_v} \int_0^u
du' g^{\alpha \beta}(u') {-} \frac{1}{2}q_5 \kappa k_v \int_0^u
du' T(u')\,, \label{Ex15}
\end{equation}
where $B(W)$ is arbitrary function of the phase $W$. When $m \neq
0$, the exact solution is described by the linear combination of
sin and cosin functions
\begin{equation}
\Phi = \frac{1}{L} \left[C_1 \cos{W} + C_2 \sin{W}  \right] \,,
\label{Ex14}
\end{equation}
where $C_1$ and $C_2$ are constants of integration. Again, the
four-gradient of the phase $W$, the wave four-vector $K^i$,
\begin{equation}
K^i \equiv \nabla^i W = \delta^i_u \left[  \frac{m^2}{2k_v} -
\frac{k_{\alpha}k_{\beta}}{2k_v} g^{\alpha \beta} - k_v
\frac{1}{2}q_5 \kappa T \right] + \delta^i_v k_v + \delta^i_2 k_2
+ \delta^i_3 k_3 \,, \label{Ex16}
\end{equation}
satisfies identically the eikonal equation $\tilde{g}^{in} K_iK_n
= m^2$, where $\tilde{g}^{in}$ is the color-acoustic metric
(\ref{Acu1}).

\section{Discussion}

According to (\ref{pp8}) and (\ref{pp9}) the quantity $L^{\prime
\prime} / L$ is negative, thus, the positive background factor
$L(u)$ is a monotonically decreasing function of the retarded
time. Starting from $L=1$ at $u=0$ this function reaches zero
value at some moment $u=u^{*}$. When $L$ tends to zero, the
functions $\Pi^2_2$ and $\Pi^3_3$ increase infinitely, and when
they pass the value $\Pi^2_2=\Pi^3_3= \pm 1$, the components of
the color metrics take zero values (see (\ref{okh3})). Are such
singularities pure coordinate ones or they have some physical
interpretation? Different aspects of the singularity, appearing in
the spacetimes, related to the pure gravitational pp-waves and
waves in the Einstein-Maxwell theory, were discussed very
intensely (see, e.g., \cite{Synge,MTW}). The standard way to
eliminate the coordinate singularity is well-known: one should
make the coordinate transformation \cite{ExactSolution}
\begin{equation}
u {=} \tilde{u} \,, \quad v {=} \tilde{v} {-} \frac{1}{2}\left[
y^2 \left(\frac{L^{\prime}}{L} {+} \beta^{\prime} \right) {+} z^2
\left(\frac{L^{\prime}}{L} {-} \beta^{\prime} \right) \right] \,,
\quad x^2 {=} y \frac{1}{L} e^{{-}\beta} \,, \quad x^3 {=} z
\frac{1}{L} e^{\beta} \,. \label{transf}
\end{equation}
Then the effective line elements (\ref{okh2}) take the form
\begin{equation}
ds^2_{(\alpha)} = 2du d\tilde{v} - dy^2 - dz^2 - 2H_{(\alpha)}
du^2 \,, \label{transf1}
\end{equation}
where
\begin{equation}
H_{(\alpha)} \equiv - \frac{1}{2}\left[ y^2 R^2_{\ u2u} + z^2
R^3_{\ u3u}\right] +  \Pi^{\alpha}_{\alpha}\,, \label{transf12}
\end{equation}
(there is no summation over $\alpha$). Thus, as in the case of
pure gravitational wave \cite{ExactSolution}, $det
(g^{ik}_{(\alpha)}) = - 1$, and the coordinate singularity is
avoided. Nevertheless, another question arises. Let us consider,
for instance, an observer the normalized velocity four-vector of
which has the $t$ component only
\begin{equation}
V^k = \delta^k_t \left[ 1 - H_{(\alpha)} \right]^{-
\frac{1}{2}}\,, \quad t = \frac{1}{\sqrt{2}} (u+\tilde{v}) \,,
\quad  x = \frac{1}{\sqrt{2}} (\tilde{v}-u) \,. \label{transf3}
\end{equation}
Such a four-vector field can not be prolonged through the surfaces
$H_{(\alpha)} = 1$, which are described by two equations ($\alpha
=2,3$) quadratic in the transverse coordinates $y$ and $z$
\begin{equation}
y^2 R^2_{\ u2u} + z^2 R^3_{\ u3u} = 2\Pi^{\alpha}_{\alpha} -2 \,,
\label{transf4}
\end{equation}
where  $R^2_{\ u2u}$ and $R^3_{\ u3u}$ are given by (\ref{ppRie1})
and (\ref{ppRie2}). These singular surfaces look like the ones,
which appear in the {\it Analog Gravity} theory
\cite{Visser1,Novello,VolovikBook}. In order to clarify the
physical sense of such (dynamic) singularity, let us consider two
particular cases of motion of the massless particle in one of the
color metrics, say, metric with $\alpha =2$.

\vspace{3mm}
\noindent {\it (i) Longitudinal motion}

\noindent Let the transverse components of the particle momentum
four-vector be equal to zero, i.e., $P^y=P^z=0$. Then one obtains
\begin{equation}
g_{ik(2)} P^iP^k =0  \ \  \rightarrow   2P^u \left(P^v -H_{(2)}P^u
\right) =0 \,. \label{obana}
\end{equation}
The first solution, $P^u=0$, relates to the uniform particle
motion in the direction $Ox$, this motion is not influenced by the
pp-wave gravity field. The second solution, $P^v=H_{(2)}P^u$
relates to the motion in the opposite direction with the
longitudinal momentum
\begin{equation}
P^x = \frac{1}{\sqrt{2}}\left(P^v-P^u \right)=
\frac{1}{\sqrt{2}}P^u \left(H_{(2)}-1 \right) \,.
\label{obana1}
\end{equation}
Such a motion is not uniform, in particular, $P^x$ vanishes, when
the particle reaches the singular surface $H_{(2)}=1$ (see
(\ref{transf4})). In other words, the particle stops when it
contacts with the surface (\ref{transf4}) and can not cross it.

\vspace{3mm}

\noindent {\it (ii) Transversal motion}

\noindent Let now $P^x=P^z=0$. Then one obtains
\begin{equation}
P^v=P^u \,, \quad (P^y)^2 = 2 (P^u)^2 \left( 1 - H_{(2)} \right)
\,. \label{obana2}
\end{equation}
Again, the particle stops at the surface $H_{(2)} = 1$, and can
not cross it. Thus, there are two singular surfaces, $H_{(2)}=1$
and $H_{(3)}=1$, which can be indicated as analogs of the
so-called {\it trapped} surfaces, discussed in \cite{Visser1}. The
color massless particles, interacting non-minimally with the
pp-wave background, can not cross these trapped surfaces.

Color-acoustic metric (\ref{Acu1}) is also accompanied by (one)
singular trapped surface. To obtain its equation we can formally
replace $H_{(\alpha)}$ by the $H_{(s)}$
\begin{equation}
H_{(s)} \equiv \frac{1}{2}\left[ (q_5-y^2) R^2_{\ u2u} + (q_5-z^2)
R^3_{\ u3u}\right] \,, \label{transf123}
\end{equation}
in the equation $H_{(s)}=1$. For the transversal motion of the
test massive scalar particle ($P^x=P^z=0$) the eikonal equation
yields
\begin{equation}
\tilde{g}_{ik} P^iP^k =m^2  \ \  \rightarrow   (P^y)^2 = 2 (P^u)^2
\left[ 1 - H_{(s)} \right] - m^2 \,. \label{obana3}
\end{equation}
This means that the particle stops at the specific surface
$H_{(s)}=1-\frac{m^2}{2(P^u)^2}$, which depends on the individual
value of $P^u$. Nevertheless, it is clear that the surface
$H_{(s)}=1$ is unreachable for all massive particles, this last
equation can be obtained in the limit $P^u \to \infty$.

In the minimal limit with pure gravitational pp-wave, when
$q_1{=}q_2{=}q_3{=}q_4{=}q_5{=}0$ and the Yang-Mills and Higgs
fields vanish, three surfaces $H_{(1)}=1$, $H_{(2)}=1$ and
$H_{(s)}=1$ coincide and form one trapped surface, describing by
equation $y^2-z^2= 2L^2 / (L^2 \beta^{\prime})^{\prime}$. Thus,
the classical results about the behaviour of the test particles in
the field of strong gravitational pp-waves can be reinterpreted
in terms of theory Analog Gravity.

\section{Conclusions}

1. Non-minimal interactions in the pp-wave
Einstein-Yang-Mills-Higgs system produce color cross-effects,
which are analogous to the magneto-electric effect in
electrodynamics. These curvature induced cross-effects result in
a new feature: the number of basic associated metrics (three for
the longitudinal and three for transversal color models) exceeds
the number of color metrics (two and two, respectively).

2. Color and color-acoustic metrics for the pp-wave
Einstein-Yang-Mills-Higgs system are presented explicitly, and
exact solutions to the Yang-Mills and Higgs equations describing
the color and color-acoustic waves are obtained for the model
with parallel fields.

3. In the framework of seven-parameter non-minimal
Einstein-Yang-Mills-Higgs model with the pp-wave symmetry six
trapped surfaces appear: two for the longitudinal color waves,
two for the transversal ones and two for color-acoustic waves
(longitudinal and transversal). The equations of these surfaces
depend on the values of five parameters $q_2$, $q_3$, $q_5$,
$Q_1$ and $Q_2$, the parameters $q_1$ and $q_4$ being hidden,
since the Ricci scalar in this model vanishes.

\section*{Acknowledgments} This
work was supported by the Deutsche Forschungsgemeinschaft through
project No. 436RUS113/487/0-5. Authors are grateful to Professor
W. Zimdahl for fruitful discussions.

\end{document}